# Volumetric Dimensioning of Strategic Stock: SOP, a step further toward a better flow control

Safa El kefi[1,2]

[1]Industrial Engineering Department, National Engineering School of Tunis, University of Tunis El Manar, Tunisia

[2]UR-OASIS, National Engineering School of Tunis, University of Tunis El Manar, Tunisia

[3]School of Systems and Enterprises, Stevens Institute of Technology, USA

Abstract

This paper is presenting a real case study. It focuses on a drug stock management problem within a Tunisian monopoly in drugs distribution: PCT. The commercial service aims to optimize the drug stock management and procurement while reducing drug shortage and cost of stock management and to have a clear vision on the estimation of the strategic stock of products. Precisely, I have investigated a set of the real depots' dataset and the Tunisian's hospitals depots of the country in particular. This analysis phase has yield to suggest some important actions, such as the drugs classification, the establishment of an adequate inventory policy for each drug family and the volumetric quantification of strategic stock.

## I-    INTRODUCTION

The planning process of the offer and request also called SOP (Sales and Operations Planning) is an essential point for the performance of the supply chain. By making appropriate analysis and involving simulations of different scenarios it makes it possible to compare alternatives by assessing their costs and other impacts avoiding waste of capacity or resources, to ensure consistency with marketing actions and actors and to ensure the customer service rate. However, activity forecasts are always affected by unforeseen constraints. From their side, the supply, storage, and distribution are also subject to uncertainties. And for production and distribution monopolies, the stock often has difficulties meeting the needs of all clients if it is not sized the way it should be. Actually, one of the objectives of this study is to propose a real-world case study of a stock management system based on optimization and help find a way to exactly approach the strategic quantity of products needed based on multiple techniques of forecasting. This aims to avoid problems of demand satisfaction in the different hospitals of Tunisia. As a consequence, the risk of understock or overstock is thus reduced, and the inventory management cost is optimized.

## II-    CONTEXT OF THE PROJECT

As part of a collaboration between ENIT, OASIS Unity of ENIT and the University of Lorraine and under the project PHC-Utica CMCU 2019, several projects are planned to identify the problems of the stock and try to solve them. A first step in this collaboration was the launch of several trainings and research to optimize the stock state. PCTs (National Pharmacy Warehouses of Tunisia) sales department deploys its efforts to meet the demand for its clients in the public hospital sector as well as in the private sector. However, due to several factors, the stock is not always enough to satisfy all orders. The sizing of the strategic stock of the hospital sector, above all, remains very difficult as evolution does not follow an estimable law. So as a first step, studying the system of Stock Management is what you need. You need to know how things are managed inside and around the storage depots, to master what you're dealing with: the inputs, the outputs of the products, the inventories, the types of orders and the planning of any action and location... Everything must be considered critically to be able to identify all the flowable failures and problems of the system.

### III- WORK
#### A- Introduction:

A good forecast surely solves this kind of problem. But the question that arises, under all these factors of uncertainty, unstable market, and demand impossible to predict in the long-term considering that the computerization of the exact history of the demand of the hospitals of the Tunisian territory turns out far to achieve, what solution is more suitable for the distribution of products between organizations without affecting the lives of PCT indirect customers and risking the over or under stock?

#### B- Drugs Classification:

We do not give enough importance to data, forgetting that information evolves over time, and the more we have exact traceability, the more powerful tools we have to predict. However, with a seasonal demand that fluctuates the forecast must be well studied and communicated, as the figure below shows, to be able to approach to a minimal error the necessary supply.

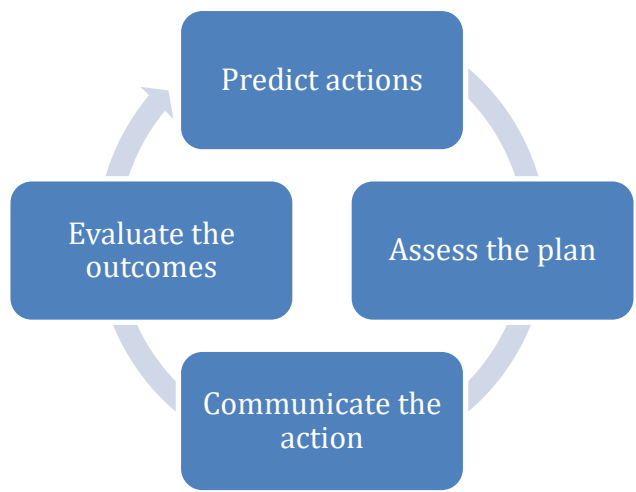

Figure 1: How forecasting should be like

Sampling and simulation are the trend of companies. Just plan the corrective actions to be good, simulate the effects on a sample and test the effects. In companies where product references are limited in number, it is easier to manage and estimate the actions to take. But the big actions have more variability when the references are more than the ten and here enters another working hypothesis which is the art of choosing its sample. Your sample needs to give information about the rest of the products. In the case of the National Pharmacy of Tunisia, the products are various and have different packages. The selection method that I used to prepare the sample of a strategic stock of the PCT was based on some selection criteria that go with the demand of the company and its management of priorities. Among the selection criteria that had to be respected: the importance of sales revenue, the quantity-price ratio of products in order to be able to put earnings into perspective, the urgency of the need for specific drugs and products compared with others. It should be mentioned that the quality of service, the speed of the orders and the satisfaction f the customers do not have enough importance when it is about monopolies companies without competing with other new entries to the market that may be hazardous to the situation and the sales. After the identification of the method comes the data filtration phase and choice of the data classification skeleton so as to be able to easily handle for the rest of the work. It can be this way considering the years of deliveries and their importance relatively to the stock.

Once the data is well ranked, the Pareto method is the best way to give the importance weight of the products according to the criteria already set. Applied the Pareto 20-80 rules, the list of products that are at the top of the list of sales and which constitute the strategic stock which, in a way, is endangered and is urgent and important to always have
available. On this sample, all proposals for corrections will be tested.

### C- Quantity of Stock Forecasting:

The idea here is to be able to estimate the stock needed for the PCT to avoid stock-out or overstocking. And, of course, improve the management of the disposal of products in the depots. After classification, comes the estimation of the exact demand for the next period based on the last years' demand. But for a better way of using the available space of storage, It is recommended to act differently to size the optimal space for the optimal quantity. And to predict the exact quantity needed, it is necessary to start from a set of hypotheses which will make that one eliminates some computational constraints or that fills missing information. In our case, these are the ones used for this case of study:

- Consumption in year N = Sales in year N-1
- Consumption Forecast = Quantity Ordered
- Consumption and sales do not obey to the fluctuations of the environment or to the seasonality and no products are withdrawn from the market for these reasons.
- The current stock will be used for sales next month for companies that need to have a stock of six months satisfied in advance, and considering these hypotheses we have for every month the Monthly Needed Quantity (M) is calculated as follows:

$$M = \frac{Livraison\ (N-1)}{12} \qquad (1)$$

and the quantity of a strategic stock (QS) is a sum of management stock equal to the consumption of the month in question with a security stock of 3 months calculated as this equation shows:

$$QS = M * 4 \quad (2)$$

So, the volumetric storage capacity must be calculated so that it satisfies the dimensions of the pallets of products. In other terms, we need to order only what is lucking considering the quantity existing already in the depots when ordering.

The quantity to be ordered (QC) must be as follows:

$$QC = QS - Stock \quad (3)$$

Here we will face two situations: Where the quantity stored exceeds the real need (over-stock) and where it is lower. It is assumed that the quantity to be ordered is zero if QC is negative and equal to QC otherwise.

### D- Volume of Stock Forecasting:

After finding the missing products in stock to reach the Stock threshold Strategy, it's time to move to the volume quantification. It is therefore necessary to take for the sample of products that have been selected, the measures of length, width and depth of the boxes of conservation of the product and calculate the already estimated need in number of boxes to an exact volume taking into consideration the carton capacity in terms of number of boxes.

### E- Seasonality effect on the forecast:

In the last part, one of the hypotheses was supposing that the demand doesn't obey to the effect of seasonality and the calculations were based on the simple moving average, which is not true, that's why in this part we're going to prove this is not true. From the list of products appearing in the sample, we chose two from two different families to test the seasonality on. If the demand is extracted in the last years and draw its representative curve the seasonality is easily noticed. Thus, the use of smoothing estimation considering the seasonality coefficients is better than the simple moving average.

What we did, is that for every year we classify the medicines based on monthly orders, so in every part, there are the sum of all the orders made by customers and distributors of the Pharmacy. Once all the data are classified according to this method, it's time to move to the selection of the products by the ABC method.

Collected the output of the classification made by the Pivot Tables in the sheets of the different years from the ABC Processing table, you move to filing products in descending order according to the quantity delivered. Due to the ponderation of the importance of the products in the operations of the company explained in the figure above, we could represent the same results.

The selection of strategic products here is only to build a model on which we will simulate our experiences for the rest of the products and drugs to improve the stock situation given the need to ensure a good rotation and availability of stock for these so-called vital products.

For volumetric sizing, the first step is taking dimensions of the cartons and boxes. Once all dimensions taken for the strategic products selected and the estimation of the strategic quantity is made, it is easy to calculate the strategic volumetric stock of the company for the next period to avoid over and under stock.

A work on a sample was first established then we have proposed a classification of all the drugs under the same method. For considering the work done, it's obvious that the optimization of the arrangement of the cartons in the pallets while at the same time maintaining the strategic stock constraint that must be in stock at the time in question. But depending on the time the information must be more and more respected and traced in order to be able to estimate the Gap between the actual demand and the offer of the company which is the main KPI to help us see the result of our actions. Once sure the solution is good enough reviewing the KPI's set, it is high time to take in consideration real actions.

The volumetric quantification of the strategic stock is more specific and precise, and it helps more in estimating the capacity while optimizing the location of products. And the result will be even more optimal if one makes use of the estimate by seasonality of the needs in medical products.

## VI- Conclusion

In this paper, we consider a real-world case study: drug inventory management in the National Pharmacy of Tunisia hospitals department. The objective of this works is to optimize the medicine management by improving the stock Management and Forecast to minimize the inventory costs.